%% file: main.tex
\RequirePackage[l2tabu, orthodox]{nag}
\documentclass[aps,pra,english,reprint,longbibliography,nofootinbib,floatfix]{revtex4-2}
\usepackage[utf8]{inputenc}
\usepackage[stretch=20]{microtype}
\usepackage{graphicx}
\usepackage[shortlabels]{enumitem}
\usepackage[draft=false, hidelinks]{hyperref}
\usepackage[textsize=tiny]{todonotes}

\input{preamble_widebar}
\input{preamble_math}
\allowdisplaybreaks
\graphicspath{{computations_and_plots/}}
\begin{document}

\title{Quench Dynamics of the Ideal Bose Polaron at Zero and Nonzero Temperatures}
\author{Moritz Drescher}
\author{Manfred Salmhofer}
\author{Tilman Enss}
\affiliation{Institut für Theoretische Physik, Universität Heidelberg,
  D-69120 Heidelberg, Germany}

\begin{abstract}
  We give a detailed account of a stationary impurity in an ideal Bose-Einstein condensate, which we call the ideal Bose polaron, at both zero and non-zero temperatures and arbitrary strength of the impurity-boson coupling.
  The time evolution is solved exactly and it is found that, surprisingly, many of the features that have been predicted for the real BEC are already present in this simpler setting and can be understood analytically therein.
  We obtain explicit formulae for the time evolution of the condensate wave function at $T=0$ and of the one-particle density matrix at $T>0$.
  For negative scattering length, the system is found to thermalize even though the dynamics are perfectly coherent.
  The time evolution and thermal values of the Tan contact are derived and compared to a recent experiment.
  We find that contrary to the Fermi polaron, the contact is not bounded at unitarity as long as a condensate exists.
  An explicit formula for the dynamical overlap at $T=0$ allows us to compute the rf spectrum which can be understood in detail by relating it to the two-body problem of one boson and the impurity.
\end{abstract}

\maketitle

\section{Introduction}
A particle immersed in a quantum mechanical medium, called polaron, is a general concept of many-body theory and a paradigmatic example of quasi-particle formation.
When the medium particles are bosonic, Bose-Einstein condensation occurs at sufficiently low temperatures.
In this case, the reaction of the bath to the local perturbation is no longer governed by individual bath particles but by a collective excitation of the condensed mode.

An instance of the polaron concept of particular relevance to date is an impurity particle immersed in an ultracold atomic gas, the realization of which has been enabled by recent experimental advances in the field.
While the Fermi polaron—the case of a fermionic environment—has been thoroughly investigated (for reviews, see \cite{Chevy2010, Massignan2014, Schmidt2018a}), the Bose polaron is a domain of active research.
It has been realized in experiments with strongly imbalanced mixtures of ultracold gases \cite{Jorgensen2016, Hu2016, Rentrop2016, Camargo2018, Skou2020, Yan2020}.

While early theoretical treatments discussed the region of a weakly coupled impurity
(see Ref.\@ \cite{Grusdt2016} for a review, Refs.\@ \cite{Kain2016, Lampo2017, Takahashi2019, Mysliwy2020} for more recent developments),
these approaches were subsequently extended towards stronger coupling
\cite{Rath2013, PenaArdila2015, PenaArdila2016, Shchadilova2016, Shchadilova2016a, Grusdt2017a, Grusdt2018, Kain2018, Drescher2019, Drescher2020, Yoshida2018, Shi2018a, Mistakidis2019, PenaArdila2019, Guenther2020}
and finite temperatures
\cite{Boudjemaa2014, Schmidt2016, Sun2017, Lampo2017, Levinsen2017, Lausch2018, Guenther2018, Dzsotjan2020}
as well as aspects of dynamics \cite{Schmidt2018, Boyanovsky2019, Nielsen2019}.
The particularly simple case of an ideal condensate and a stationary impurity—since a mobile impurity would induce effective interactions—has been frequently treated as a limiting case to theories developed for the interacting Bose gas
\cite{Rath2013, Volosniev2015, Christensen2015, Shchadilova2016, Shi2018a, Yoshida2018, Drescher2019, Guenther2020},
but not received an extensive treatment on its own, with methods specifically devoted to its solution.

In this paper, we fill this gap by focusing entirely on this problem, to which we refer as the ideal Bose polaron.
Considering this simple setting has several benefits:
First of all, the comparison of ideal and real polaron (through experiments or interacting theories) allows to understand better the role of boson interactions.
This is most interesting for a strongly coupled impurity, where the BEC is significantly deformed and even weak boson interactions can have an important effect.
When relaxation effects are observed in a system, it is not always clear whether they originate from many-body effects, dephasing or incoherent dynamics.
In an ideal BEC, the situation is simpler because there is no incoherence, i.e.\@ an initial coherent or product state keeps this form as time evolves.
This leads to insights on the nature of peak broadening in rf spectra and of thermalization.
The simplicity of the problem allows to understand effects analytically that were previously predicted by interacting theories:
In particular, there is a close relation to the two-body problem of one boson and the impurity and many-body effects can be related to the two-body modes.
Last but not least, the results we obtain are exact and require little
numerical effort even at nonzero temperature.

As one central result of this article, we solve the time evolution after a sudden switch-on of the impurity interactions at $t=0$ (a \textit{quench}) exactly and show that the condensate wave function evolves as $(ħ = 1)$
\begin{multline}
  ϕ(\vec r, t\smallrel{>}0) =
  \sqrt{\nB} - \sqrt{\nB} \frac{\aIB}{r} e^{-\mB r^2 / 2it} \Biggl[
       \erfcx{\biggl( \frac{r}{2}\sqrt\frac{2\mB}{it} \biggr)} \\
       - \erfcx{\biggl( \frac{r}{2} \sqrt\frac{2\mB}{it} - \frac{1}{\aIB} \sqrt\frac{it}{2\mB} \biggr)}
     \Biggr]
  \label{eq:wave_function}
\end{multline}
where $\erfcx z = e^{z^2} (1 - \erf z)$ is the scaled complementary error function, $\nB$ the boson density, $\mB$ the boson mass and $\aIB$ the impurity-boson scattering length.
Similar formulae hold for the excited states and allow us to compute the reduced density matrices at nonzero temperature.
We find that thermalization occurs for $a < 0$ by dephasing but not for $a > 0$ because low-lying bound states cannot be populated without incoherences.
From the density matrix, we obtain the Bose polaron's Tan contact at $T > 0$ for the first time and compare to results from a recent experiment \cite{Yan2020}, for which we give a new interpretation.
We then derive the explicit formula for the dynamical overlap at $T=0$, compute the exact rf spectrum and discuss how it emerges analytically.

The paper is organized as follows.
Sec.\@ \ref{sec:definitions} introduces the model.
Sec.\@ \ref{sec:dynamics} discusses the dynamics and the asymptotics for long times and the question of thermalization.
Sec.\@ \ref{sec:results} presents physical observables: The Tan contact is discussed, including the question of unitarity limitation and comparison to experiment. The rf spectrum at $T=0$ is computed and related to the two-body problem.

The zero-temperature results of this article are based on Ref.\@ \cite{Drescher2020a}.

\section{Model} \label{sec:definitions}
The Hamiltonian of an ideal BEC in presence of a stationary impurity is the one-particle operator
\begin{align*}
  \HmanyIB &= ∫ \dd{^3𝐫} a^{\dagger}_𝐫 \HIB_𝐫 a^{}_𝐫 \\ %∑_i \HIB_{𝐫_i} \\ %-∑_i \frac{Δ_{\vec r_i}}{2\mB} + ∑_i \VIB(\vec r_i),
  \HIB_𝐫 &= -\frac{\mathrm{Δ}_𝐫}{2\mB} + \VIB(𝐫)
\end{align*}
where the impurity is situated at the origin, $a_𝐫$ annihilates a boson at position $𝐫$ and $\VIB$ is the impurity-boson interaction potential with scattering length $a$.
In typical cold-atom experiments, the potential range is the shortest length scale by far and the interactions can be modelled by an effective potential of zero range, the Fermi pseudo-potential
\begin{equation*}
  \VIB ψ(\vec r) = \frac{2π\aIB}{\mB} δ^3(\vec r)∂_r (rψ(𝐫)).
\end{equation*}
It can have negative or positive scattering length and has, accordingly, zero or one bound states.
Furthermore, it acts only on s-wave states.

The resulting eigenstates of $\HIB$ are the continuum states $\mode_k$, parameterized by the wave number $k > 0$, the bound state $\mode_\bound$ if $a > 0$ and the zero-mode $\mode_0$, which we separate from the continuum because of its macroscopic occupation by the condensate:
\begin{subequations} \label{eqs:eigenstates}
\begin{align}
  \mode_0 &= 1 - \frac{\aIB}{r}  \\
  \mode_\bound &= \frac{\exp(-r/\aIB )} {r \sqrt{2π\aIB }}      && \text{if }\aIB  > 0 \\
  \mode_k &= \frac{\sin(kr) - \aIB k\cos(kr)} {rπ\sqrt{2(1 + \aIB ^2k^2)}} && \text{for }k > 0
\end{align}
\end{subequations}
with energies $E_0 = 0$, $E_\bound = - 1 \divslash 2\mB \aIB^2$ and $E_k = k^2 \divslash 2\mB$.
The normalization is such that $∫_{ℝ^3} \abs{\mode_\bound}^2 = 1$, $∫_{ℝ^3} \conj{\mode_q} \mode_k = δ^1(k - q)$ for $k, q > 0$ and $\mode_0 \rightarrow  1$ as $r \rightarrow  ∞$.

We are interested in the dynamics of a BEC which is initially in its ground state or thermal state and which starts to interact with the impurity at $t=0$.

\subsubsection{$T=0$}
At zero temperature, the Schrödinger equation is solved by a product state ansatz
$Ψ(𝐫_1 … 𝐫_{\NB}, t) = ϕ(𝐫_1, t) ⋯ ϕ(𝐫_{\NB}, t)$
where $\NB$ is the number of bosons.
The \textit{condensate wave function} $ϕ$ is normalized to $∫_{\vol} \abs{ϕ}^2 = \NB$, where $\vol$ is the system volume, which allows for convergence in the thermodynamic limit.
The dynamics of $ϕ$ are determined by the two-body Schrödinger equation of one boson and the impurity, $i∂_t ϕ = \HIB ϕ$, with an initially flat BEC, $ϕ(t\smallrel{=}0) = \sqrt n$, of density $n = N/\vol$.

\subsubsection{$T > 0$}
At nonzero temperature, the condensate is best described in terms of the $m$-particle density matrices \cite{Lieb2002, Erdos2009, Yngvason2014} $γ_m(𝐱_1 … 𝐱_m, 𝐲_1 … 𝐲_m) = \bigl< a_{𝐲_1}^{\dagger} \cdots a_{𝐲_m}^{\dagger} a_{𝐱_1}^{} \cdots a_{𝐱_m}^{} \bigr>$
where $\left< … \right> = \Tr ρ … $ denotes the expectation value with respect to the density matrix $ρ$.
Their dynamics are usually determined by the BBGKY hierarchy \cite{Huang1987} but since $\HmanyIB$ is a one-particle operator, the hierarchy decouples into
\begin{equation*}
  i∂_t γ_m = \Bigl[ ∑_{i=1}^m \HIB_i, γ_m \Bigr].
\end{equation*}
We may therefore restrict to $γ_1$, which gives access to the expectation values of one-boson operators, such as the ground-state energy, the bosonic density profiles, the Tan contact between one boson and the impurity, etc.\footnote{In fact, $γ_1$ gives access to all $γ_m$: their initial states are given by symmetrized tensor products of $γ_1$ and this structure is preserved by time evolution w.r.t.\@ a one-particle Hamiltonian.}
The initial state is given in terms of the modes of the free two-body Hamiltonian $\Hfree$.
By separating the s-wave modes $\freemode_0$, $\freemode_k$ from the rest, we write
\begin{align*}
  γ_1 ={}& γ_1^\textrm{s-wave} + γ_1^\textrm{other} \\
  γ_1^\textrm{s-wave}(t\smallrel{=}0) ={}& n_0 \projector{\freemode_0} \\
    & {} + \intpos \dd q \frac{1}{z^{-1} e^{β E_q} - 1} \projector{\freemode_q }.
\end{align*}
Here, $n_0 = n(1 - (T/\Tc)^{3/2})$ is the condensate density with the critical temperature $\Tc$ and $z$ is the fugacity, determined by $z = 1$ if $T ≤ \Tc$ and $\Li_{3/2}(z) = (\Tc/T)^{3/2} \Li_{3/2}(1)$ otherwise ($\Li$ is the polylogarithm).

\section{Exact Dynamics} \label{sec:dynamics}

Even though the condensate wave function obeys the same Schrödinger equation as in the two-body problem, there is an important difference:
As $\NB, \vol \rightarrow  ∞$, the condensate wave function $ϕ$ becomes non-normalizable, which leads to qualitatively different dynamics.
Also note that if $\VIB$ allows for a bound state, the energy is not bounded below because in absence of boson-boson interactions, an arbitrary number of bosons can enter the bound state.

\subsection{$\mathbfit{T} = \mathbf{0}$}
At zero temperature, the condensate is fully condensed in the zero-mode $\freemode_0 = 1$ and its time evolution is given by $ϕ(\vec r, t) = \sqrt{\nB} \freemode_0(t)$ where
\begin{equation*}
  \freemode_0(t) = e^{-it\HIB}\freemode_0.
\end{equation*}

\subsubsection{Expansion in Two-Body Eigenstates}
To solve the time evolution of $\freemode_0$, we expand it in the interacting eigenstates \eqref{eqs:eigenstates}:
\begin{multline} \label{eq:decomposition}
  \freemode_0(t) = α_{0,0} \mode_0 + α_{0,\bound} \mode_\bound e^{-itE_\bound} + \intpos \dd k α_{0,k} \mode_k e^{-itE_k}
\end{multline}
with coefficients $α$ given by%
\footnote{Derivation: Since $\cramped{\mode_0} \rightarrow  1$ as $r \rightarrow  ∞$ while the other modes decay, it is clear that $α_{0,0} = 1$.
For the bound state and continuum, it is sufficient to project the remaining part $\freemode_0 - α_{0,0} \mode_0 = a / r$ onto $\mode_\bound$ and $\mode_k$, which is easily accomplished with the aid of formula \eqref{eq:psi_alternative} below and integral \eqref{eq:integral_distributional} from the appendix.}
(we will always include the bound state in general expressions and set $α_\bound = 0$ if $\aIB < 0$)
\begin{subequations} \label{eq:coefficients_0}
\begin{align}
  α_{0,0}      &= 1 \\
  α_{0,\bound} &= θ(\aIB) 2\aIB \sqrt{2π\aIB} \\
  α_{0,k}      &= \frac{2\aIB \sqrt{2}}{k\sqrt{1 + \aIB^2 k^2}}. \label{eq:coefficient_0_k}
\end{align}
\end{subequations}
The continuum integral in \eqref{eq:decomposition} is solved in the next section, which leads to \eqref{eq:wave_function}.
Three qualitative features can be directly observed from the decomposition \eqref{eq:decomposition}.
\begin{enumerate}[label=$(\roman*)$]
  \item $α_{0,\bound}$ is finite.
  Thus the ground state, in which an infinite number of bosons is bound to the impurity, is never reached with a quench but instead the system remains dynamically stable.
  \item Quantum mechanical systems with multiple energy levels give rise to stable oscillations between these levels while oscillations involving the continuum dephase for long times.
  In \eqref{eq:decomposition}, however, if the continuum dephases, convergence towards $\mode_0$ occurs for $\aIB < 0$ while for $\aIB > 0$, oscillations between $\mode_0$ and $\mode_\bound$ remain.
  Thus, the zero mode plays the role of an additional bound state.
  We have investigated the phenomenon of stable oscillations in our previous works \cite{Drescher2019, Drescher2020}, see also Refs.\@ \cite{Kain2018, Grusdt2017a, Shchadilova2016}.
  \item There are infinitely many particles in the continuum since $n \intpos \dd k \absq{α_{0,k}}$ diverges as $𝒪(n\vol^{1/3})$,
  even though this is a vanishing fraction of all particles $N=n\vol$.
  This effect is related to the vanishing of the dynamical overlap of the ground states of $\Hmanyfree$ and $\HmanyIB$ for $a < 0$, which has been termed bosonic orthogonality catastrophe in Refs.\@ \cite{Yoshida2018, Mistakidis2019, Guenther2020}.
\end{enumerate}

\paragraph*{Global Phase}
In a finite-volume system, the lowest scattering state has not an energy of precisely zero but of $2π\aIB \divslash \mB\vol$ to leading order in $\vol$.
Even though this tends to zero in the thermodynamic limit, it cannot be neglected because the mode is occupied by a macroscopic number $N_0$ of bosons.
This leads to a total finite energy $2π\aIB n \divslash \mB$ which can be taken into account by a global phase factor:
\begin{equation*}
  Ψ(\vec r_1 … \vec r_{\NB}) = e^{-it2π\aIB n \divslash \mB} ϕ(\vec r_1) ⋯ ϕ(\vec r_{\NB}).
\end{equation*}
Even though this energy equals the mean-field result which is approached for weak coupling for the real Bose polaron \cite{Rath2013}, it remains exact for the ideal polaron at arbitrary coupling strengths.
Recall that this energy, also referred to as stationary energy, corresponds to the ground state only for $\aIB < 0$ while for $\aIB > 0$, the ground state energy is $-\infty$.

\subsubsection{Solving the Continuum Integral} \label{sec:solve_integral_0}
To obtain formula \eqref{eq:wave_function}, we have to solve the continuum integral in \eqref{eq:decomposition}.
For this calculation, we set $2m = 1$; the complete result can be recovered by scaling $t$ appropriately.
We make use of the following alternative form of the continuum modes:
\begin{equation} \label{eq:psi_alternative}
   \mode_k = \frac{-\Re e^{ikr}(\aIB k + i)} {rπ\sqrt{2(1 + \aIB ^2k^2)}}.
\end{equation}
The general types of integrals that occur in the derivation are computed in Appendix \ref{app:integrals}.
With this, the calculation of the time evolved condensate wave function proceeds as follows:
\begin{alignat*}{2}
  \MoveEqLeft[0] \intpos α_{0,k} \mode_k(r) e^{-itk^2} \dd k \\
  & \begin{annotate}
      After inserting the definitions \eqref{eq:coefficient_0_k} and \eqref{eq:psi_alternative}, $(ak + i)$ can be cancelled.
      The integrand is symmetric in $k$, so we can replace $\cramped{\intpos \rightarrow  \frac{1}{2}\intinf}$. For later use, introduce an $\Im{t} < 0$.
    \end{annotate} \\
  &= \frac{-\aIB}{rπ} \lim_{\Im{t} ↗ 0} \intinf e^{-ik^2t} \Re{}
     \frac{e^{ikr}}{k(\aIB k-i)} \dd k\\
  & \begin{annotate}
      The imaginary part of the fraction is anti-symmetric in $k$, so the `Re' can be omitted. This requires using a principal value integral since the imaginary part has a pole at 0. Apply a partial fraction decomposition.
    \end{annotate} \\
  &= \frac{-\aIB}{rπ} \lim_{\Im{t} ↗ 0} \pint e^{-ik^2t + ikr}
     \left(\frac{i}{k} - \frac{\aIB i}{\aIB k-i} \right) \dd k \\
  & \begin{annotate}
      Use integrals \eqref{eq:integral_pv}, \eqref{eq:integral_with_i}.
    \end{annotate} \\
  &= \frac{\aIB}{r} \Biggl[
       \erf{\biggl( \frac{r}{2\sqrt{it}} \biggr)} \\
       & &\mathllap{ {}- e^{it / \aIB^2 - r/\aIB}
         \biggl( \sgn{\aIB } -
           \erf{\biggl( \frac{\sqrt{it}}{\aIB } - \frac{r}{2\sqrt{it}} \biggr)}
     \biggr) \Biggr] } .
\end{alignat*}
Upon re-insertion of $2m$ and in combination with the other terms in \eqref{eq:decomposition}, this yields the exact time evolution of the condensate wave function \eqref{eq:wave_function}.
%\begin{multline*}
%  \freemode_0(𝐫,t) =
%  1 + \frac{a}{r} \Biggl[
%      -1 + \erf{\biggl( \frac{r}{2\sqrt{it}} \biggr)} \\
%         + e^{it / \aIB^2 - r/\aIB}
%      \biggl( 1 +
%             \erf{\biggl( \frac{\sqrt{it}}{a} - \frac{r}{2\sqrt{it}} \biggr)}
%           \biggr)
%    \Biggr].
%\end{multline*}

\subsubsection{Asymptotics}
For long times, the wave function \eqref{eq:wave_function} becomes
\begin{subequations}
\begin{align}
  \MoveEqLeft[0] ϕ(\vec r, t) \notag \\
  &\underset{t\rightarrow ∞}{≃} \sqrt{\nB} - \sqrt{\nB} \frac{\aIB}{r}
                  \left(1 - e^{-\frac{r}{\aIB}} \erfcx \biggl( - \frac{1}{\aIB}\sqrt\frac{it}{2\mB} \biggr) \right)
                    \label{eq:asymptotics1} \\
  &\underset{\hphantom{t\rightarrow ∞}}{≃} \sqrt{\nB} \begin{cases}
                      1 - \frac{\aIB}{r}
                          & \text{if } \aIB ≤ 0 \\
                      1 - \frac{\aIB}{r} \left( 1 - 2 e^{-\frac{r}{\aIB} + \frac{it}{2\mB {\aIB}^2}} \right)
                          & \text{if } \aIB > 0.
                    \end{cases}
                    \label{eq:asymptotics2}
\end{align}
\end{subequations}
The convergence from \eqref{eq:asymptotics1} to \eqref{eq:asymptotics2} is logarithmically slow in amplitude but qualitatively, the features of \eqref{eq:asymptotics2} are present already after short times.
As expected, convergence to $\mode_0$ is observed for $\aIB < 0$ while oscillations between $\mode_0$ and $\mode_\bound$ remain for $\aIB > 0$: indeed, \eqref{eq:asymptotics2} is simply eq.\@ \eqref{eq:decomposition} with the explicit values of $α_{0,0}$ and $α_{0,\bound}$ but without the continuum part, which has dephased.

For $\aIB > 0$, the asymptotic wave function periodically has a zero whenever the time evolution factor is $-1$.
This leads to a halo of depletion of the condensate density at a certain distance $r_0$ around the impurity.
This had been observed before for an interacting Bose gas \cite{Drescher2019} and we are now able to obtain $r_0$ analytically:
from $1 = \frac{\aIB}{r_0} \left( 1 + 2 e^{-r_0/\aIB} \right)$, we obtain a universal halo radius of $r_0 \divslash \aIB ≈ 1.45$.
Comparison with \cite{Drescher2019} indicates that this is not significantly modified by the presence of a boson-boson interaction.

\subsection{$\mathbfit{T} > \mathbf{0}$}
At nonzero temperature, we must compute the time evolution of the single-particle density matrix $i∂_t γ_1 = [\HIB, γ_1]$.
Since the potential acts only on s-wave states, the result can be written as
\begin{align}
  γ_1(t) ={}& γ_1^\textrm{s-wave}(t) + γ_1^\textrm{other}(0) \notag \\
  γ_1^\textrm{s-wave}(t)
    ={}& n_0 \projector{\freemode_0(t)} \notag \\
    & {}+ \intpos \dd q \frac{1}{z^{-1} e^{βE_q} - 1} \projector{\freemode_q(t)}. \label{eq:gamma}
\end{align}
The time evolution of the noncondensed modes $\freemode_q$ can be solved analogously to $\freemode_0$.
We obtain the decomposition
\begin{equation} \label{eq:decomposition_q}
  \freemode_q(t) = α_{q,\bound} \mode_\bound e^{-itE_\bound} + \intpos \dd k \, α_{q,k} \mode_k e^{-itE_k}
\end{equation}
with coefficients ($𝒫$ denotes the principal value)
\begin{subequations} \label{eq:coefficients_q}
\begin{align}
  α_{q,\bound} &= \frac{2qa^{3/2}}{\sqrt{π}(1 + a^2 q^2)} \\
  α_{q,k} &= \frac{1}{π\sqrt{1 + a²k²}} \left(πδ(k-q) + 𝒫 \frac{2akq}{k²-q²} \right).
\end{align}
\end{subequations}
The continuum integral is a bit more involved than for the zero-mode but it can be solved with the same techniques.
The derivation can be found in appendix \ref{app:solve_integral_q} and leads to an explicit formula for the time evolution of the noncondensed modes:
\begin{align}
   \freemode_q(t) ={}& e^{-itq²\divslash 2m} \frac{\sin(qr)}{rπ\sqrt{2}}
   + \frac{aq e^{-m r²\divslash 2it}}{rπ\sqrt 2 (1 + a²q²)} \notag \\
   & \begin{alignedat}{2}
      {} \times \Biggl(
          && \frac{iaq - 1}{2} \erfcx \biggl(\frac{r}{2}\sqrt\frac{2m}{it}
              + iq\sqrt\frac{it}{2m} \biggr) \\
          &&    - \frac{iaq + 1}{2} \erfcx \biggl(\frac{r}{2}\sqrt\frac{2m}{it}
                - iq\sqrt\frac{it}{2m} \biggr) \\
          &&  + \erfcx \biggl(\frac{r}{2}\sqrt\frac{2m}{it}
                - \frac{1}{a} \sqrt\frac{it}{2m}\biggr) & \Biggr).
  \end{alignedat}
  \label{eq:time_evolution_q}
\end{align}
The remaining integral in \eqref{eq:gamma} can easily be solved numerically.

\begin{figure*}
  \centering
  \includegraphics{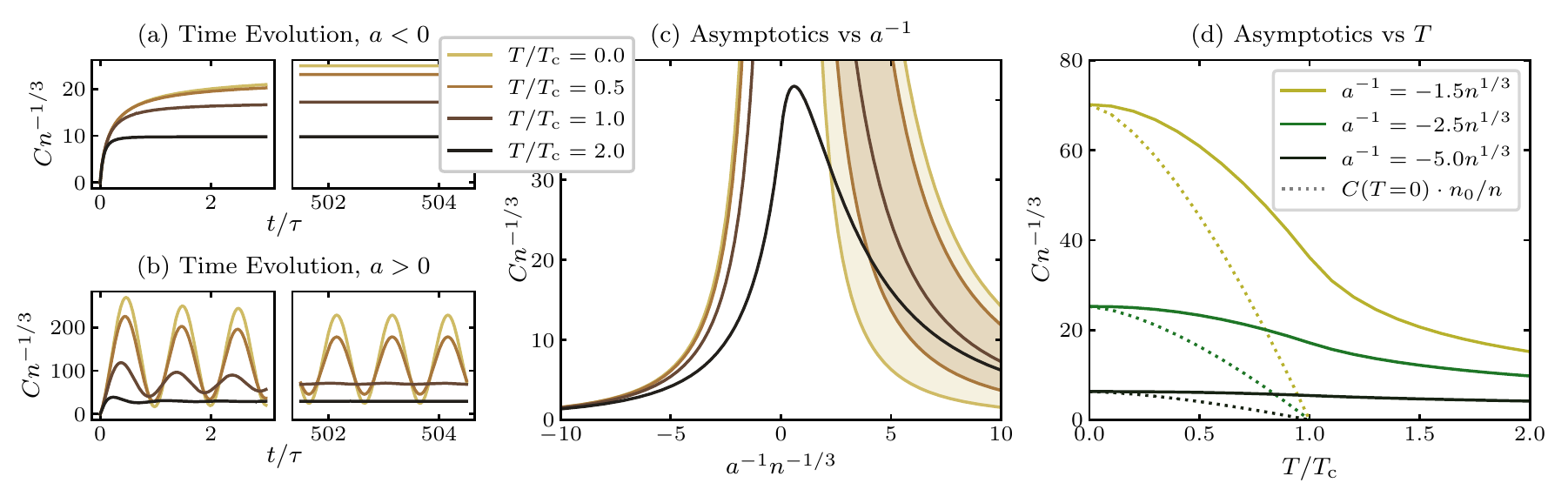}
  \caption{Impurity-induced Tan contact.
  \textbf{(a, b)}~Time evolution for $a^{-1} = \pm 2.5n^{1/3}$ at various temperatures.
  Thermalization occurs for $a < 0$ while for $a > 0$, stable oscillations remain whenever $T < \Tc$.
  Time is measured in units of $τ = 2\mB n^{-2/3}$.
  \textbf{(c)}~Asymptotic values vs.\@ scattering length.
  For $a>0$, the contact keeps oscillating within the shaded regions.
  %At $T=0$, the contact is given by $16π^2 a^2 n$ for $a<0$, while for $a > 0$, it oscillates between this value and its nine-fold.
  For growing temperatures, the amplitude of oscillations decreases proportional to the condensate density.
  Below $\Tc$, the contact diverges at unitarity.
  This is in contrast to the Fermi polaron, where Pauli blocking causes a limitation even in an ideal gas.
  \textbf{(d)}~Temperature dependence of the thermalized values for $a < 0$.
  In comparison the contribution from the condensate (dotted).
  At weak coupling, the thermal part compensates for the decrease in condensate density and the contact is almost constant even well above $\Tc$.
  At stronger coupling, the compensation is only partly, resulting in a decay $C(T)$ which is strongest around $\Tc$.
  }
  \label{fig:contact}
\end{figure*}

\subsubsection*{Asymptotics and Thermalization}
In presence of boson interactions, one would expect the system to finally reach its thermal state.
To investigate if thermalization can occur even for the coherent dynamics of the ideal setting, we look at the long-time limit of $\freemode_q$ and find
\begin{equation*}
  \freemode_q(t) \underset{t\rightarrow ∞}{≃} \frac{\abs{1+iaq}}{1+iaq} e^{-itE_q} \mode_q
  + α_{q,\bound} e^{-itE_\bound} \mode_\bound,
\end{equation*}
similarly as for the zero-mode (the $t$-independent phase factor in front could as well be absorbed in the definition of $\mode_q$).
For $a < 0$, convergence to the interacting modes results and $γ_1$—and with it the entire system—thermalizes.
For $a > 0$, on the other hand, the bound state leads once again to oscillations.
Under the continuum integral, these oscillations dephase and vanish for observables that have a continuous momentum space representation.
Thus, the asymptotic $γ_1$ can effectively be written as
\begin{multline*}
  γ_1^\textrm{s-wave}(t, T) \underset{t\rightarrow ∞}{≃} \frac{n_0}{n} γ_1^\textrm{s-wave}(t; T \smallrel = 0) \\
  {} +   \intpos \dd q \frac{1}{z^{-1} e^{βE_q} - 1} \Bigl( \projector{\mode_q} + \absq{α_{q, \bound}} \projector{\mode_\bound} \Bigr).
\end{multline*}
For $T < \Tc$, oscillations with an amplitude proportional to $n_0 / n$ remain, while for $T > \Tc$, the reduced density matrix converges.
Still, it does not thermalize for $a > 0$ since the energy is not bounded below and no thermal state exists.

\section{Observables} \label{sec:results}
We now turn to the discussion of physical observables that can be directly inferred from the condensate wave function \eqref{eq:wave_function} at $T=0$ or obtained by numerically solving the integral in \eqref{eq:gamma} for $T>0$.

\subsection{Tan Contact}
The Tan contact was first introduced as a thermodynamic quantity in an interacting quantum gas \cite{Tan2008}.
In the polaron problem, the impurity-induced contact $C = \bigl< \lim_{r\rightarrow 0} (4πr)^2 a_𝐫^{\dagger} a^{}_𝐫 \bigr>$ quantifies the short-distance singularity of the host atom density around the impurity.
It has been measured for the Bose polaron via its relation to $ω^{-3/2}$ tails in the rf spectrum \cite{Yan2020}.

\subsubsection{$T = 0$}
From the analytical formula for the condensate wave function \eqref{eq:wave_function}, we can read off the time evolution of the contact $C = \abs{4π \lim_{r\rightarrow 0} rϕ(r)}^2$ at $T = 0$:
\begin{align} \label{eq:contact}
  C(t) = 16π^2a^2n \abs*{1 - \erfcx \biggl(-\frac{1}{a} \sqrt\frac{it}{2m} \biggr)}^2.
\end{align}

Figs.\@ \ref{fig:contact}(a, b) show the time evolution for two scattering lengths with opposite sign.
The characteristic convergent or oscillating behavior of the wave function discussed in the last section is clearly visible.
For long times, the asymptotic curve
\begin{equation} \label{eq:contact_asymptotic}
  C(t) \underset{t\rightarrow ∞}{≃} 16π^2a^2n \begin{cases}
     1 & \text{if } a ≤ 0 \\
     \abs{1 - 2e^{-itE_\bound}}^2 & \text{if } a ≥ 0
   \end{cases}
\end{equation}
is reached.
Fig.\@ \ref{fig:contact}(c) displays the dependence of the final values on the scattering length.

In particular, the contact diverges as the resonance $1/a = 0$ is approached and an infinite number of bath particles are attracted.
This behavior is different from the Fermi polaron: there, Pauli blocking prevents unbounded growth.
This effect is known as unitarity limitation and leads to a universal unitary value of the contact that is only weakly affected by the intra-species interactions.
The question whether such a universal value exists for the bosonic case as well has so far been discussed only with approximate methods that contain an \textit{a priori} limitation of the number of excitations from the condensate and therefore cannot exhibit a divergence.
Here, we have shown that in absence of Pauli blocking no unitarity limitation exists.
Instead, the contact keeps growing linearly in time as \eqref{eq:contact} takes the simple form%
\footnote{Linear growth of the contact has also been predicted when the intra-species interactions in a Bose gas are quenched to unitarity \cite{Sykes2014}.}
$C(t) = 32πnt\divslash m$ for $a^{-1} = 0$.
Only the presence of a boson repulsion can prevent unbounded growth in a real condensate.
But then, the value of the contact at unitarity should depend on the parameters of this boson interaction and not be universal:
In particular, we expect that it continues to grow as $\aBB \rightarrow  0$ at unitarity.

\subsubsection{$T > 0$}
Only the s-wave part of $γ_1$ has a nonzero contribution to the contact, which is thus given by
\begin{align}
  C(t, T) ={}& \frac{n_0}{n} C(t, T\smallrel{=}0) \notag \\
    & + \lim_{r\rightarrow 0} (4πr)² \intpos \dd q \frac{1}{z^{-1} e^{βE_q} - 1} \absq{\freemode_q(r, t)}. \label{eq:contact_T}
\end{align}
Its time evolution and asymptotic behavior for various temperatures are shown in Fig.\@ \ref{fig:contact}.
For $a < 0$, the contact decreases slowly with growing temperature but retains a significant part of its $T\smallrel{=}0$ value even above $\Tc$.
For $a > 0$, the contribution from the zero-mode keeps oscillating but the thermal part converges as discussed before in more generality.

At unitarity, the contact continues to diverge below $\Tc$ while it remains finite above $\Tc$:
the unitarity limitation is present in the uncondensed Bose gas but not in a condensate.
The asymptotic value above $\Tc$ is given by $C = 4\sqrt{2m π/β} \Li_{1/2}(z)$ for $a^{-1} = 0$ (a similar formula has recently been derived for the Fermi polaron \cite{Liu2020}).
Below $\Tc$, the diverging contribution of the condensate is added but curiously, even the thermal part diverges since $\Li_{1/2}(1) = ∞$.

According to \eqref{eq:contact_T}, the contact can be split into a part originating from the condensate and a thermal part.
Fig.\@ \ref{fig:contact}(d) shows this decomposition for the thermalized values at $a < 0$.
For weak coupling, no temperature dependence is visible: Here, the thermal part compensates almost completely for the reduced condensate density as the temperature is raised.
At stronger coupling, a decrease with temperature is observed, but the thermal part still has a significant contribution.

\begin{figure*}
  \centering
  \includegraphics{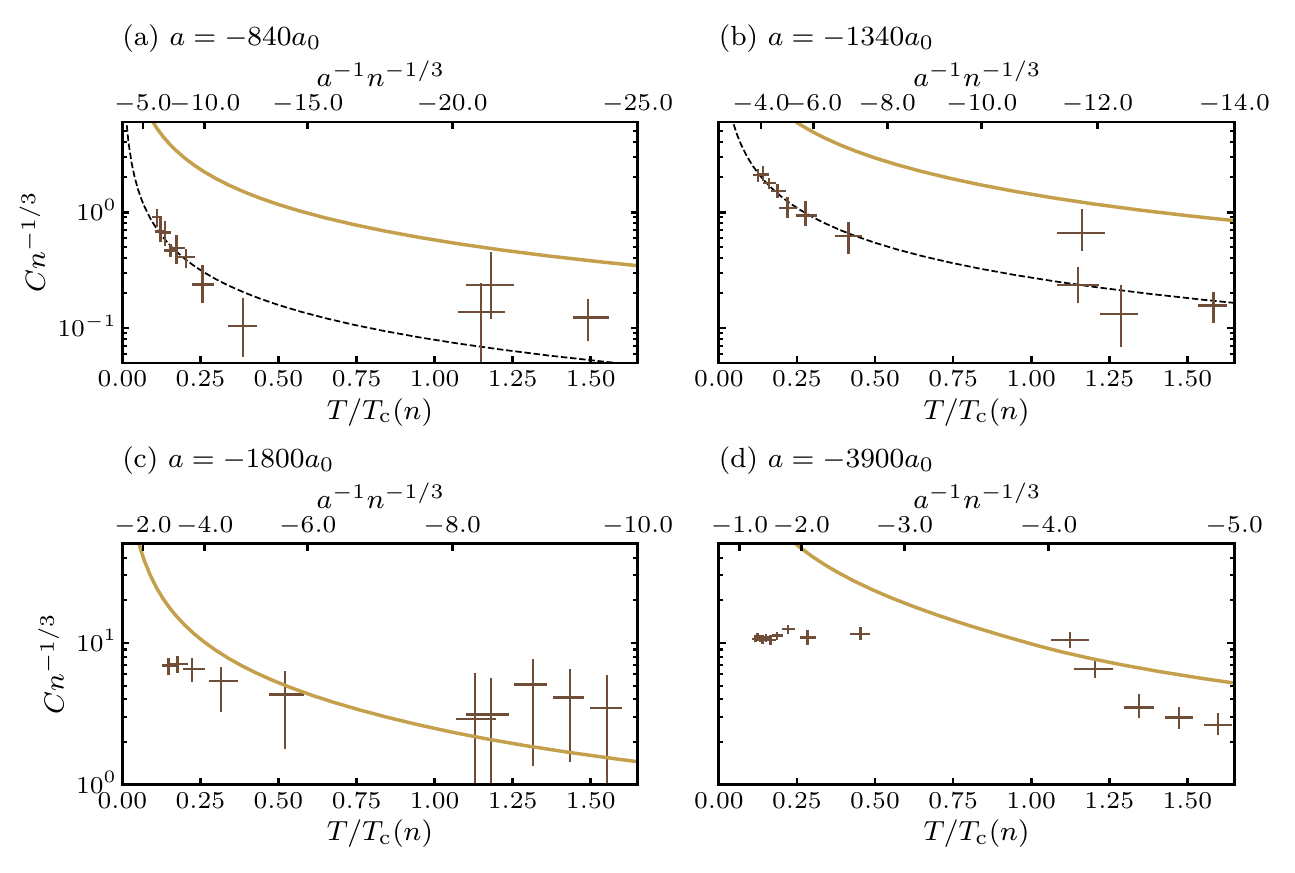}
  \caption{%
  %\textbf{(b–e)}
  Comparison of the Tan contact of the ideal Bose polaron to the recent experiment of Yan \textit{et al.\@} \cite{Yan2020}.
  At fixed absolute temperature $T ≈ \SI{130}{nK}$ and scattering length (see captions, $a_0$ is the Bohr radius), the density is varied by evaluating different trap regions, leading to a simultaneous change in $T/\Tc ∼ n^{-2/3}$ and $an^{1/3}$.
  \textbf{Crosses} are the experimental data points with error bars \cite{Yan2020a}.
  The leftmost data points correspond to the trap center and the density decreases towards the right.
  In (a, b), the \textbf{dashed lines} are $n^{2/3}$ fits to the experiment (i.e.\@ $C ∼ n$).
  From the excellent agreement we conclude that temperature and boson repulsion have no effect here and that the variation is due to the change in $an^{1/3}$.
  The \textbf{solid lines} are our exact results for the ideal polaron.
  The behavior in (a, b) agrees remarkably well up to constant factors of about 7 and 4.5.
  The origin of this prefactor is not yet clarified.
  In (d), the experimental values remain constant below $\Tc$.
  Here, the strong-coupling regime is entered, in which the limiting effect of the boson interaction $\aBB$ becomes important.
  This breaks the $n^{2/3}$ scaling and there is no longer qualitative agreement with the ideal polaron.
  Note that also the boson coupling scales as $\aBB n^{1/3}$ in local units and thus has a stronger effect towards the left.
  }
  \label{fig:compare_experiment}
\end{figure*}

\subsubsection{Comparison to Experiment}
The contact was recently measured experimentally \cite{Yan2020}.
In this work, the contact was extracted from the high-frequency tail of rf spectra of a trapped imbalanced gas mixture at fixed temperature and scattering lengths.
By evaluating different trap regions with different densities, a variation of $T / \Tc$ was achieved, as the critical temperature depends on $n$ via
\begin{equation*}
   k_{\text{\textsmaller B}} \Tc(n) = \frac{2π}{m} \left( \frac{n}{\Li_{3/2}(1)} \right)^{2/3}.
\end{equation*}
At the same time, the local length unit $n^{-1/3}$ changes, leading to an effective variation of the impurity coupling $an^{1/3}$ and of the boson repulsion $\aBB n^{1/3}$.
Thus, while $T$, $a$ and $\aBB$ are fixed in absolute units, they effectively vary in units of $n$ which are called local units in Ref.\@ \cite{Yan2020}.

In Fig.\@ \ref{fig:compare_experiment} we compare the experimental data of the four measurement series to the asymptotic (thermalized) values of eq.\@ \eqref{eq:contact_T}, subject to the same parallel variation in $\Tc$ and $an^{1/3}$ with density.
The comparison allows us to discuss the similarities and differences between ideal and real polaron, i.e.\@ the effects of Boson repulsion and impurity mass.
The expectation is that the relatively low experimental value of $\aBB$ has no significant effect for a weakly coupled impurity but that it may get important near unitarity where it limits the number of particles attracted by the impurity.
A finite impurity mass on the other hand should have a weaker effect at strong coupling where it is enhanced to a larger effective mass.

\subsubsection*{Weak Coupling (a, b)}
The experimental data seem to imply, at first sight, a strong decay of the contact with growing temperature.
However, comparison with Fig.\@ \ref{fig:contact}(d) shows that temperature should have vanishing effect for the coupling strengths in Fig.\@ \ref{fig:compare_experiment}(a, b).
We thus conjecture that the decrease originates from the effective change in scattering length.
To verify this, note that the non-influence of $T$ manifests itself in the scaling $C ∼ na^2$ \eqref{eq:contact_asymptotic} (see also Ref.\@ \cite{PenaArdila2015}) of the contact, which leads to $Cn^{-1/3} ∼ n^{2/3} a^2 ∼ (T/\Tc)^{-1}$ for the experimental scaling $\Tc ∼ n^{2/3}$.
%For these parameters, comparison with Fig.\@ \ref{fig:contact}(d) and the NLGPT\todo{not there} shows that neither temperature nor boson repulsion should have a significant influence on the contact.
%The experimental values decrease, however, when the density is lowered (left to right in the plots) and we conclude that this originates from the effective change in scattering length.
%To test this further, the non-influence of $T$ and $\aBB$ manifests itself in the scaling $C ∼ na^2$ \eqref{eq:contact_asymptotic} of the contact, which leads to $Cn^{-1/3} ∼ n^{2/3} a^2 ∼ (T/\Tc)^{-1}$ for the experimental scaling $\Tc ∼ n^{2/3}$.
Such a $1/x$-fit is shown as dashed line in the figures and agrees remarkably well.
%for (a, b), while for (c), the situation is less clear due to the large error bars.

Our exact result (solid line) shows the same $1/x$ behaviour but deviates by constant factors.
It is not plausible that this difference originates from the difference between ideal and real Bose gas for a number of reasons:
\begin{enumerate}[(i)]
  \item The difference is highest for the lowest scattering length (a), but the limiting effect of the boson repulsion should be most important for strong coupling.
  \item Comparison with approaches that include boson interactions \cite{Drescher2020, Guenther2020} indicates that the latter has almost no effect at these coupling strengths.
  \item The boson coupling $\aBB n^{1/3}$ effectively varies with density too, which would break the $1/x$ scaling.
\end{enumerate}
That instead the finite impurity mass should cause the difference would explain why the factor is smaller for the stronger coupling (b) where the effective mass is larger.
However, given that the impurities are already heavier than the bosons in the experiment, such a large effect would be surprising.
The origin is thus not yet clear and investigating the role of the mass ratio on the Tan contact might be an interesting topic for future work.

\subsubsection*{Strong Coupling (d)}
%Near unitarity, the experimental values are approximately constant near the trap center and theory and experiment no longer show the same qualitative behavior.
%Here, real-world effects that are not included in the ideal setting become important, as they limit the number of bosons the impurity can attract.
%A boson repulsion in the gas would certainly have this effect; another candidate is the finite impurity density that allows only for a limited number of attracted bosons per impurity.
Near unitarity, agreement between ideal and real Bose polaron can no longer be expected because the former predicts a divergence of the number of attracted particles, which is limited by the repulsion between Bosons in the real gas.
Consequently, the experimental and theoretical results in (d) no longer show the same qualitative behavior.
Also a finite density of impurities might have a limiting effect in the experiment as it allows only for a finite number of attracted bosons per impurity.

For a simple estimate of the magnitude of these effects, let us assume that each impurity has at its disposition a limited volume of radius $R$ in which it may disturb the BEC.
The zero-mode then turns into $\mode_0 = \frac{1 - a/r}{1 - a/R}$, which becomes $\mode_0 = R/r$ at unitarity and leads to a contact limited by $C ∼ nR^2$.
\begin{itemize}
\item For the effect of the boson repulsion, we assume that any perturbation of the BEC is limited in size to the magnitude of the healing length $ξ = 1 \divslash \sqrt{8π \aBB n_0}$.
Setting thus $R ∼ ξ$ leads to $Cn^{-1/3} ∼ \frac{n}{n_0} \frac{1}{\aBB n^{1/3}}$.
In the experiment, $\aBB^{-1} n^{-1/3} ≈ 100$ at the trap center and both $n/n_0$ and $\aBB^{-1} n^{-1/3}$ grow towards the outer regions.
\item For the impurity concentration, we set $R ∼ n_I^{-1/3}$ and obtain $Cn^{-1/3} ∼ (n / n_I)^{2/3}$.
In the experiment, $(n / n_I)^{2/3} ≈ 45$ is constant under the assumption of similar cloud shapes.
\end{itemize}
That the measured contact is constant might lead to the conclusion that the latter effect is dominant.
However, even for this measurement series, $a^2n^{2/3} ≈ 0.7$ is much lower than $(n / n_I)^{2/3}$ and the impurity concentration should have no effect.
Rather, we believe that the boson repulsion is augmented by a prefactor, which depends on the potential \emph{shape} and is larger for a van-der-Waals potential than for a soft one.
This expectation is motivated by the findings of Ref.\@ \cite{Drescher2020} but whether the suggested scaling as $\aBB^{-1}n^{-1/3}$ is valid for arbitrary potential shapes remains to be seen.

The constant experimental values of $C$ near unitarity may be caused by three effects that compete as the density is lowered from the trap center to the outer regions:
\begin{itemize}
  \item The decrease in $an^{1/3}$ leads to a decrease of the contact in the same units.
  \item The increase in $T/\Tc$ also leads to a decrease, cf.\@ Fig.\@ \ref{fig:contact}(d).
  \item The decrease in $\aBB n^{1/3}$ on the other hand leads to an increase.
\end{itemize}
In total, these effects seem to compensate for each other remarkably well.

\subsubsection*{Summary}
To summarize, our results for the ideal Bose polaron allow us to interpret the observed density dependence of the experimental data at weak coupling:
comparison with Fig.\@ \ref{fig:contact}(d) and 1/x fits shows that the temperature has vanishing effect at these coupling strengths and that, consequently, the variation originates from the effective change in coupling strength.
We have found a surprising deviation between ideal and real polaron at weak coupling that may motivate research on the influence of the impurity mass on the Tan contact.
Near unitarity, on the other hand, the lower experimental values are well-interpretable by boson repulsions in a real gas that limit the number of bosons gathering around the impurity.

\subsection{Dynamical Overlap and RF Spectrum at $\mathbfit{T} = \mathbf{0}$}
While the Tan contact between bosons and impurity is a one-particle operator of the form $∫ \dd{𝐱}
\dd{𝐲} O_{𝐱, 𝐲} a_𝐱^{\dagger} a^{}_𝐲$, experiments give also access to many-body observables which contain an arbitrary number of creators and annihilators.
The two most relevant are the dynamical overlap
\begin{equation*}
  S(t) = \bigl< e^{it\Hmanyfree} e^{-it\HmanyIB} \bigr>
\end{equation*}
(the expectation value is with respect to the initial state), which can be measured by Ramsey interferometry \cite{Goold2011, Cetina2016, Skou2020}, and its Fourier transform \cite{Schmidt2018a}, the rf spectrum
\begin{equation} \label{eq:A_functional}
  A(ω) = \bigl< ∫ \dd E δ(E - \Hmanyfree) δ(ω - E + \HmanyIB) \bigr>,
\end{equation}
which is measured by spectroscopy \cite{Hu2016, Jorgensen2016, Yan2020}.

They have been computed theoretically both at zero \cite{Rath2013, Shashi2014, Shchadilova2016, Jorgensen2016} and non-zero \cite{Field2020, Dzsotjan2020} temperatures and Refs.\@ \cite{Shchadilova2016, Jorgensen2016, Dzsotjan2020} have compared to experiment with good agreement.
We will now derive these quantities exactly for the ideal polaron at zero temperature and show that the features known from the real polaron are already present---even though some of them were believed to originate from incoherences in the dynamics, which are absent in our case.

\subsubsection{Dynamical Overlap}
From the decomposition \eqref{eq:decomposition}, we obtain the dynamical overlap (switching from product states to coherent states):
\begin{alignat}{2}
  S(t) &= \braket{Ψ(0) | Ψ(t)} \divslash \braket{Ψ | Ψ} \notag\\
       &= \exp\biggl(-it2π\aIB n \divslash \mB + n \abs{α_{0,\bound}}^2 \bigl( e^{-itE_\bound} - 1 \bigr) \notag\\
       && \mathllap{ {}+ n \intpos \dd k \, \abs{α_{0,k}}^2 \bigl( e^{-itE_k} - 1 \bigr) \biggr)}. \label{eq:dynamical_overlap_decomposed}
\end{alignat}
The solution of the continuum integral is similar as in  section \ref{sec:solve_integral_0}. We set $2m=n=1$ for the calculation:
\begin{alignat*}{2}
  \MoveEqLeft[0] \intpos \bigl(e^{-ik^2 t} - 1 \bigr) \abs{α_{0,k}}^2 \dd k \\
  & \begin{annotate}
      Again, replace $\intpos \rightarrow  \frac{1}{2} \intinf$ and perform a partial fraction decomposition.
    \end{annotate} \\
  &= 4\aIB^2 \biggl( ∫_ℝ \frac{e^{-ik^2t} - 1}{k^2} \dd k {}
                    -\aIB ^2 ∫_ℝ \frac{e^{-ik^2t} - 1}{1 + \aIB^2 k^2} \, \dd k \biggr) \\
  & \begin{annotate}
      First integral, call it $I(t)$: $\frac{∂}{∂t} I(t) = -i \intinf e^{-ik^2t} = -i \sqrt{\frac{π}{it}}$, thus $I(t) = -2\sqrt{iπt}$.
      Second integral: \eqref{eq:integral_quadratic_denom} from the appendix.
    \end{annotate} \\
  &= 4\aIB^2 \Biggl( -2\sqrt{iπt} - \aIB π e^{it/\aIB^2} \biggl(\sgn \aIB  - \erf \frac{\sqrt{it}}{\aIB } \biggr) + \abs{\aIB} π \Biggr) & .
\end{alignat*}
Upon re-insertion of $2m$ and $n$ and combining with the contributions of the zero-mode and bound state, we obtain the exact formula for the dynamical overlap:
\begin{multline}
  S(t) = \exp \Biggl[ -it2π\aIB n \divslash \mB
    + 4π\aIB ^3n \Biggl( \erfcx\biggl(-\frac{1}{\aIB}\sqrt\frac{it}{2\mB} \biggr)
                       \\ {} - 1 - \frac{2}{\aIB } \sqrt\frac{it}{2\mB π} \Biggr) \Biggr].
  \label{eq:dynamical_overlap}
\end{multline}
At short times, it shows oscillations for $a > 0$ similarly to the contact, but for long times $S$ decays to zero \cite{Shchadilova2016, Yoshida2018, Mistakidis2019, Guenther2020}.
This is remarkable given that the dynamics are perfectly coherent and that the condensate wave function continues to oscillate.
This difference in behaviour of contact and dynamical overlap originates from the former being a one-particle and the latter a many-particle observable.

In the two limiting cases of weak and of unitary coupling $a^{\pm 1} \rightarrow  0$, formula \eqref{eq:dynamical_overlap} simplifies to a stretched exponential with a noninteger power of time:
\begin{subequations} \label{eq:dynamical_overlap_limit_cases}
\begin{align}
  S(t) &\xrightarrow[a\rightarrow 0]{\hphantom{a\rightarrow ∞}} \exp\left(-\frac{2πan}{m}it - 8a^2 n \sqrt{π} \Bigl( \frac{it}{2m} \Bigr)^{1/2} \right) \label{eq:dynamical_overlap_weak} \\
  S(t) &\xrightarrow[a\rightarrow ∞]{} \exp\left(\frac{16\sqrt{π}n}{3} \Bigl(\frac{i t}{2m}\Bigr)^{3/2} \right). \label{eq:dynamical_overlap_strong}
\end{align}
\end{subequations}
Note in particular that the unitary limit $a^{-1} \rightarrow  0$ is well-behaved despite the divergence of the zero-mode energy $2πan/m$ because the latter is cancelled by the continuum contribution (c.f.\@ the continuum shift in the rf spectrum discussed below).

\subsubsection{RF Spectrum}
\begin{figure}
  \includegraphics{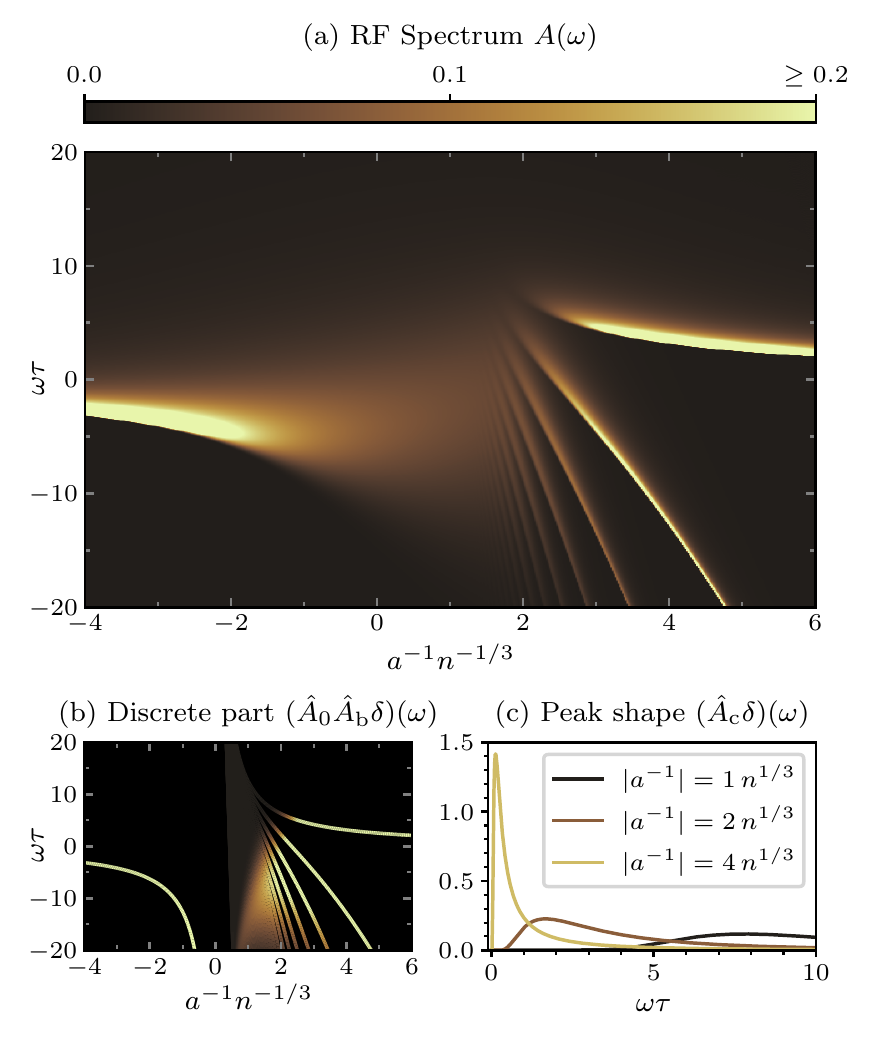}
  \caption{\textbf{(a)} Numerically exact rf spectrum of the ideal Bose polaron.
  It can be understood as originating from discrete peaks \textbf{(b)}: The main peak corresponds to the energy of the zero-mode $2π\aIB n \divslash \mB$ and for $\aIB > 0$, additional lines appear at distances of the two-body binding energy.
  Their weights follow a Poisson distribution with more lines having appreciable weight at strong coupling.
  The discrete peaks are broadened and shifted by the continuum, giving rise to the peak shape in \textbf{(c)}.
  Since continuum states have positive energy, the lines in (b) are situated at the lower bounds of those in (a).
  %The broadening and shifting is strongest close to the resonance and leads to a smooth crossover.
  }
  \label{fig:rf_spectrum}
\end{figure}

The explicit formula \eqref{eq:dynamical_overlap} gives us access to the exact rf spectrum via a single numerical Fourier transform,
\begin{equation*}
  A(ω) = \frac{1}{π} \Re \intpos e^{itω} S(t) \dd t.
\end{equation*}
The result is shown in Fig.\@ \ref{fig:rf_spectrum}(a) for a range of scattering lengths.
It is similar to the result previously obtained from Bogoliubov theory for the real polaron \cite{Shchadilova2016}.
However, since Bogoliubov theory has a dynamical instability at strong coupling \cite{Grusdt2017a, Drescher2019}, it was not clear before whether the results remain correct near unitarity.
The agreement with our exact results shows that the instability does not affect the rf spectrum despite the fact that it leads to divergences in other observables \cite{Drescher2019}, such that we can confirm Ref.\@ \cite{Shchadilova2016}, with the exception of one difference:
In our Fig.\@ \ref{fig:rf_spectrum}, the rf peaks can be seen to be quite sharply cut off towards lower energies for weak coupling.
Also, from \eqref{eq:A_functional} and $\Hmanyfree ≥ 0$, one can see that $A(ω)$ is strictly zero when $ω$ is below the ground state energy of $\HmanyIB$ for $a < 0$.
In Ref.\@ \cite{Shchadilova2016}, however, the peaks appear to be broadened in both directions and centered around the ground state energy.
We do not believe that this difference originates from a difference between ideal and real BEC nor from the approximations involved in Ref.\@ \cite{Shchadilova2016} but instead from the higher resolution that our explicit formula for $S(t)$ allows to obtain with little numerical effort.

\subsubsection*{Analytical Discussion}
The emergence of the rf spectrum from the three constituents of the two-body spectrum—the zero mode, the bound state and the continuum—can be understood in detail.
For this, we split up \eqref{eq:dynamical_overlap_decomposed} into three corresponding factors
$S_0$, $S_\bound$ and $S_\cont$,
interpret them as multiplication operators acting on the constant function $1$ and Fourier transform each of these operators independently:
\begin{align*}
  S(t) &= (S_0 S_\bound S_\cont 1)(t) \\
  A(ω) &= (\Aop_0 \Aop_{\bound} \Aop_{\cont} δ) (ω),
\end{align*}
where $\Aop_{\dots} = ℱ^{-1} S_{\dots} ℱ$ and $ℱ$ is the Fourier transform operator, scaled as $(ℱf)(t) = ∫_{ℝ} e^{-itω} f(ω) \dd{ω}$.
The result is a $δ$-peak centered around zero which is subsequently modified by the operators $\Aop_0$, $\Aop_\bound$ and $\Aop_\cont$.
They can be expressed in terms of the shift operator $σ_E$ defined by $σ_E f(ω) = f(ω - E)$
and their effects can be discussed independently.%
\footnote{There is a related probabilistic description: We may regard $A$ as probability density of a random variable which is an infinite sum of random variables corresponding to the different modes.
From this, one can infer that $A$ is an \textit{infinitely divisible distribution}.
In accordance with this, $\ln S$ satisfies the \textit{Lévy-Khinchin formula}, i.e., it is of the form
\begin{equation*}
  \ln S(t) = -\frac{σ^2}{2} t^2 - ibt + ∫ (e^{-itE} - 1 + itE 𝟙_{\abs{E} < 1}) \mathop{ν(dE)}
\end{equation*}
without Gaussian part, $σ^2 = 0$, and with the term $E𝟙_{\abs{E} < 1}$ being $ν$-integrable such that it can be absorbed in $b$ (see \eqref{eq:dynamical_overlap_decomposed}).
While the zero-mode is responsible for the deterministic shift $b$,
the bound state and continuum correspond to the singular and absolutely continuous parts of the characteristic measure $ν$. }

\paragraph{Zero-Mode}
For the zero-mode, we have
\begin{align*}
  S_0 &= e^{-it2π\aIB n \divslash \mB} \\
  \Aop_0 &= σ_{2π \aIB n \divslash \mB}.
\end{align*}
It thus shifts the delta function by $2π\aIB n \divslash \mB$, leading to the main peaks in Fig.\@ \ref{fig:rf_spectrum}.
At stronger coupling, they are no longer visible due to the effect of the continuum discussed below.

\paragraph{Bound State} For $a > 0$, the bound state has the effect
\begin{align*}
  S_\bound &= \exp \left(n\abs{α_{0,\bound}}^2 \bigl(e^{-itE_\bound } - 1\bigr) \right) \\
  \Aop_\bound &= \exp \left(n\abs{α_{0,\bound}}^2 (σ_{E_\bound} - 1) \right).
\end{align*}
Fig.\@ \ref{fig:rf_spectrum}(b) shows the purely discrete part of the spectrum $(\Aop_0 \Aop_\bound δ)(ω)$.
Expanding the exponential yields
\begin{equation} \label{eq:bound_state_shifts}
  \Aop_\bound = ∑_{j=0}^∞ \frac{(n\abs{α_{0,\bound}}^2)^j}{j!}e^{-n\abs{α_{0,\bound}}^2} σ_{jE_\bound}.
\end{equation}
We thus obtain shifts by multiples $jE_\bound$ of the two-body binding energy as discussed in Ref.\@ \cite{Shchadilova2016}:
They correspond to many-body eigenstates of $j$ bosons in the bound state and $N-j$ in the zero-energy state.
The weights of these peaks follow a Poisson distribution%
\footnote{This is immediately plausible: Every boson has a probability of $\absq{α_{0, \bound}}/\vol$ to be found in the bound state and the Poisson distribution emerges through the law of rare events.}
with mean value $\left<j\right> = n\abs{α_{0,\bound}}^2 = 8π\aIB^3n$.
For weak coupling, this value is small, meaning that the bound state is occupied at most once.
For stronger coupling, multiply occupied bound states gain weight and additional lines appear in the spectrum.
Close to unitarity, the lines get denser and the broadening by the continuum stronger, such that a smooth crossover to the attractive side is observed \cite{Shchadilova2016}.

\paragraph{Continuum} For the continuum, we obtain
\begin{align*}
  S_\cont &= \exp \biggl(n\intpos \dd k \, \abs{α_{0,k}}^2 \bigl( e^{-itE_k} - 1 \bigr) \biggr) \\
  \Aop_\cont &= \exp \biggl(n\intpos \dd k \, \abs{α_{0,k}}^2 ( σ_{E_k} - 1 ) \biggr).
\end{align*}
The influence of the continuum on a single peak is shown in Fig.\@ \ref{fig:rf_spectrum}(c); it is independent of the sign of $a$.
It broadens the peaks, but only towards positive $ω$ because the continuum states have exclusively positive energies.
At the same time, a shift occurs which cancels the mean effect of zero-mode and bound states:
This is required by the sum rule $∫ ω A(ω) \dd{ω} = \braket{Ψ(0) | \HmanyIB | Ψ(0)} = 0$.
In the limiting cases $a^{\pm 1} \rightarrow  0$, the power-law exponent $∼ t^α$ in \eqref{eq:dynamical_overlap_limit_cases} means that $A$ is an $α$-stable distribution.%
\footnote{The stable distributions are a narrower class than the infinitely divisible distributions and are fully characterised by a stability parameter $α ∈ (0, 2]$, a skewness parameter $β ∈ [-1,1]$ as well as a shift and a scale parameter. }
Specifically for weak coupling, we obtain the Lévy distribution (stability $α = 1/2$, skewness $β = 1$)
\begin{equation} \label{eq:A_cont_levy}
  (\Aop_\cont δ)(ω) \xrightarrow[a\rightarrow 0]{} θ(ω) \frac{4\aIB^2 n}{\sqrt{2m} \, ω^{3/2}} \exp\left(-\frac{8π\aIB^4 n^2}{mω} \right).
\end{equation}
%For weak coupling, the peak shape can be estimated:
%According to \eqref{eq:dynamical_overlap_weak}%
%\footnote{In both cases \eqref{eq:dynamical_overlap_limit_cases} the power-law exponent $∼ t^α$ means that $A$ is an $α$-stable distribution, which is a narrower class than the infinitely divisible distributions.
%Specifically, for $a \rightarrow  0$ $A$ approaches the Lévy distribution \eqref{eq:A_cont_levy} while for $a^{-1} = 0$, a Holtsmark distribution is obtained.}
%we have $S_\cont ≈ \exp(-8\aIB^2n \sqrt{iπt\divslash 2m})$
%which has the Fourier transform
%\begin{equation} \label{eq:A_cont_levy}
%  \Aop_\cont δ(ω) ≈ θ(ω) \frac{4\aIB^2 n}{\sqrt{2m} \, ω^{3/2}} \exp\left(-\frac{8π\aIB^4 n^2}{mω} \right).
%\end{equation}
It exhibits the $ω^{-3/2}$ tail characteristic for contact interactions.
In Fig.\@ \ref{fig:rf_spectrum}(c), the narrowest peak is of this form.
At unitarity on the other hand, $A(ω)$ is the stable distribution with stability $α = 3/2$ and skewness $β=1$ and has a representation in terms of Whittaker functions \cite{Zolotarev1954, Zaliapin2005}.

%It has been conjectured that the peak broadening in the rf spectrum is a sign of incoherence in the dynamics \cite{Yan2020}.
%This possibility is ruled out by our analysis of perfectly coherent dynamics which nonetheless show broad peaks near unitarity.
Interestingly, the time evolution of the system is perfectly coherent but leads nonetheless to broad peaks near unitarity, which arise from the coupling of the impurity to the continuum.
It should also be noted that while the rf spectrum does contain information about the many-body Hamiltonian $\HmanyIB$, it also depends on the noninteracting Hamiltonian $\Hmanyfree$ and the initial state of the system.
For example, the positions of the discrete peaks are direct features of $\HmanyIB$, but their weights as well as the continuum effects relate to $\Hmanyfree$.
In particular, the strong influence of the continuum and the broad peaks near unitarity are a consequence of the fact that the difference between $\HmanyIB$ and $\Hmanyfree$ is most important here.

\section{Conclusion}
We have treated the problem of a stationary impurity in an ideal BEC and derived exact results for the condensate wave function at zero temperature and the reduced density matrix at nonzero temperature.
For attractive coupling, the system is found to thermalize by dephasing despite the lack of incoherence in the dynamics because the two-body spectrum is purely continuous.
On the repulsive side oscillations remain, which are proportional in amplitude to the condensate fraction.

We used our results to compute the Tan contact, which grows unboundedly at unitarity, contrary to the behavior for the Fermi polaron.
This effect vanishes above the critical temperature and will be suppressed by boson interactions in a real gas.
The dependence of the contact on temperature is small for weak coupling and increases towards unitarity.
We have compared our results to a recent experiment and interpreted the varying contact along different trap regions as being due to an effective change in coupling strength.

The rf spectrum at zero temperature was shown to have the same qualitative features as they have been predicted for the interacting Bose gas \cite{Shchadilova2016} and can be understood very well in the ideal setting from its relations to the two-body problem.

For future work, the absence of unitarity limitation in an ideal BEC reported here naturally leads to the question how the boson interaction in a real gas would prevent unlimited attraction of bosons by the impurity.
We have estimated the scaling to be $\aBB^{-1}n^{-1/3}$ which has yet to be confirmed by explicit calculations.
Also the shape of the interaction potential can be expected to play an important role \cite{Drescher2020}.
Furthermore, the influence of the impurity mass on the contact promises to be interesting:
Important questions are if the contact of a mobile impurity at unitarity in an ideal Bose gas would continue to diverge and if the mass ratio can explain the difference reported here between the exact results for the ideal polaron and the experimental results for the real polaron.

\textit{Note added.}---After completion of this manuscript, a study of universal aspects of the Bose polaron appeared \cite{massignan2020} that also finds the contact to diverge at unitarity for $\aBB\to0$.
A more recent work on the dynamical formation of the polaron derives the same expression \eqref{eq:dynamical_overlap_strong} for the unitary dynamical overlap \cite{PenaArdila2020}.

\begin{acknowledgments}
  We acknowledge interesting discussions with M. Zwierlein, Z. Z. Yan, M. Weidemüller, B. Tran, M. Gerken, E. Lippi, M. Rautenberg and B. Zhu.
  This work is supported by the Deutsche Forschungsgemeinschaft (DFG, German Research Foundation) under project-ID 273811115 (SFB1225 ISOQUANT) and under Germany's Excellence Strategy EXC2181/1-390900948 (the Heidelberg STRUCTURES Excellence Cluster).
\end{acknowledgments}

\appendix

\section{Continuum Integral for Noncondensed Modes} \label{app:solve_integral_q}
By similar means as in section \ref{sec:solve_integral_0} for the zero-mode, we solve the integral in the decomposition \eqref{eq:decomposition_q} to obtain the time evolution of the continuum modes \eqref{eq:time_evolution_q}.
Once again, we set $2m = 1$ and make use of the integrals in appendix \ref{app:integrals}.
\begin{alignat*}{2}
  \MoveEqLeft[0] \intpos \dd k e^{-itk²} α_{q,k} \mode_k(r) \\
  &= \frac{-1}{rπ²\sqrt2}  \intpos \dd k e^{-itk²} \Re \frac{e^{ikr}}{ak-i} \biggl( πδ(k-q)  \\
  &&\mathllap{ {}+ 𝒫 \frac{2akq}{k²-q²} \biggr) }\\
  & \begin{annotate}
    Evaluate the $δ$-function.
    For the other term, replace $\cramped{\intpos \rightarrow  \frac{1}{2}\intinf}$ and introduce $\Im{t} < 0$ as before.
    Then apply a partial fraction decomposition.
  \end{annotate} \\
  &= \frac{-1}{rπ²\sqrt2} \lim_{\Im t ↗ 0} \Biggl[ π e^{-itq²} \Re \frac{e^{iqr}}{aq-i} \\
  && \mathllap{ {} + \frac{aq}{2(1 + a²q²)} \pint \dd k e^{-itk² + ikr} }\\
  && \mathllap{ {} ⋅ \left( \frac{aq+i}{k-q} - \frac{aq-i}{k+q} - \frac{2ia}{ak-i} \right) } & \Biggr] \\
  & \begin{annotate}
      Apply equations \eqref{eq:integral_with_i}, \eqref{eq:integral_pv}.
  \end{annotate} \\
  &= \frac{-1}{rπ\sqrt2 (1 + a²q²)} \Biggl[ e^{-itq²} \Re e^{iqr}(aq+i) \\
  &&\mathllap{ {}+ \frac{iaq(aq+i)}{2} e^{-itq² + iqr} \erf \biggl(\frac{r}{2\sqrt{it}} + iq\sqrt{it} \biggr) }\\
  &&\mathllap{ {}- \frac{iaq(aq-i)}{2} e^{-itq² - iqr} \erf \biggl(\frac{r}{2\sqrt{it}} - iq\sqrt{it} \biggr) }\\
  &&\mathllap{ {}+ aq e^{it/a² - r/a} \left(\sgn a + \erf \biggl(\frac{r}{2\sqrt{it}} - \frac{\sqrt{it}}{a} \biggr) \right) }& \Biggr] \\
  & \begin{annotate}
    Some rearrangements.
  \end{annotate} \\
  &= e^{-itq²} \frac{\sin(qr)}{rπ\sqrt2} + \frac{aq e^{-r²/4it}}{2rπ\sqrt2 (1 + a²q²)} \Biggl[ \\
  &&\mathllap{  i(aq+i) \erfcx \biggl(\frac{r}{2\sqrt{it}} + iq\sqrt{it} \biggr) }\\
  &&\mathllap{ {}- i(aq-i) \erfcx \biggl(\frac{r}{2\sqrt{it}} - iq\sqrt{it} \biggr) }\\
  &&\mathllap{ {}+ 2 e^{it/a² - r/a + r²/4it} \left(-\sgn a - \erf \biggl(\frac{r}{2\sqrt{it}} - \frac{\sqrt{it}}{a} \biggr) \right) } & \Biggr].
\end{alignat*}
The part of the bound state is
\begin{equation*}
    e^{-itE_\bound} α_{q,\bound} ψ_\bound = 2θ(a) e^{it/a² - r/a}
   \frac{qa}{rπ\sqrt2 (1 + a²q²)}.
\end{equation*}
Together, this leads to eq.\@ \eqref{eq:time_evolution_q}.

\section{Derivation of Required Integrals} \label{app:integrals}

For the solution of the continuum integrals and the projections occurring throughout the paper, the following integrals are needed.
\begin{subequations} \label{eq:integrals}
\begin{alignat}{2}
  \intinf \frac{e^{-αk^2}}{k^2 + γ^2} \dd k
    &= \frac{π}{γ} e^{αγ^2} \bigl(\sgn(\Re γ) - \erf(γ\sqrt α) \bigr)
    \label{eq:integral_quadratic_denom} \\
  \intinf \frac{e^{-α k^2 + β k}} {k - iγ} \dd k
    &= iπ e^{αγ^2+iβγ} \biggl(\sgn(\Re γ)  \notag\\
    & &\mathllap{ {}- \erf\Bigl(\frac{iβ}{2\sqrt α} + γ\sqrt α \Bigr) \biggr) \hspace{1.5em} }
    \label{eq:integral_with_i} \\
  \pint \frac{e^{-α k^2+β k}} {k - q} \dd k
    &= iπ e^{-αq^2 + βq} \erf \Bigl(iq\sqrt α - \frac{iβ}{2\sqrt α} \Bigr)
    \label{eq:integral_pv} \\
  \intpos e^{iqr} \dd{r} &= π δ(q) + 𝒫 \frac i q
     \vphantom{\frac{e^{q^2}}{q}} \label{eq:integral_distributional}
\end{alignat}
\end{subequations}
where $α,β,γ ∈ ℂ$, $q ∈ ℝ$ with $\Re α > 0$, $\Re γ ≠ 0$
and (d) holds in the sense of distributions.
%The principal value `$\pv{}$' in $\pv{\frac 1k}$ is left implicit in applications.

\subsection*{Proof}
In the following it is always easy to find dominating functions to show holomorphy of integrals and exchangeability of differentiation and integration.
\begin{enumerate}[(a)]
\item Denote by $I_1(α, γ)$ the left-hand side of \eqref{eq:integral_quadratic_denom}.
We have
\begin{align*}
  \frac{∂}{∂α} e^{-αγ^2} I_1(α,γ) &= -e^{-αγ^2} ∫_ℝ e^{-α k^2} \dd k \\
  &= -\sqrt{\frac{π}{α}} e^{-αγ^2}.
\end{align*}
Consequently,
\begin{align*}
  I_1(α, γ) &= -e^{αγ^2} ∫^α \sqrt{\frac{π}{α'}} e^{-α'γ^2} \dd{α'} \\
  & \begin{annotate}[5cm]
      $s = γ \sqrt{α'}$
    \end{annotate} \\
  &= -\sqrt π e^{αγ^2} ∫^{γ\sqrt{α}} e^{-s^2} \frac{2\,\dd s}{γ} \\
  &= -\frac{π}{γ} e^{αγ^2} \bigl(\erf(γ\sqrt α) + \textit{const}(γ) \bigr)
\end{align*}
With $I_1(+∞, γ) = 0$, we obtain
\begin{equation*}
  I_1 = \frac{π}{γ} e^{αγ^2} \left(\sgn(\Re γ) - \erf(γ\sqrt α) \right) .
\end{equation*}
\item Let $I_2(α,β,γ)$ be the integral in \eqref{eq:integral_with_i}.
Then
\begin{align*}
  \frac{∂}{∂β} e^{-iγβ} I_2 &= e^{-iγβ} ∫_ℝ e^{-α k^2 + β k} \dd k \\
  &= \sqrt{\frac{π}{α}} e^{\frac{β^2}{4α}-iγβ}.
\end{align*}
Therefore,
\begin{alignat*}{2}
I_2 &= e^{iβγ} ∫^β \sqrt{\frac{π}{α}} e^{\frac{{β'}^2}{4α} - iγβ'} \dd{β'} \\
  & \begin{annotate}[5cm]
      $s = \frac{iβ'}{2\sqrt α} + γ\sqrt α$
    \end{annotate} \\
  &= \sqrt{\frac{π}{α}} e^{iβγ} ∫^{\frac{iβ}{2\sqrt{α}} + γ\sqrt{α}} e^{-s^2 + γ^2α} \frac{2\sqrt{α}}{i} \, \dd s \\
  &= -iπ e^{iβγ + γ^2α} \biggl( \erf\Bigl(\frac{iβ}{2\sqrt{α}} + γ\sqrt{α} \Bigr) \\
  && \mathllap{ {}+ \textit{const}(α, γ) \biggr) }.
\end{alignat*}
To evaluate the constant, compute $I_2(α, 0, γ)$:
\begin{align*}
  I_2(α, 0, γ) &= ∫_ℝ e^{-α k^2} \frac{k + iγ}{k^2 + γ^2} \dd k \\
  &= iγ ∫_ℝ \frac{e^{-α k^2}}{k^2 + γ^2} \\
  &= iγ I_1 \\
  &= iπ e^{αγ^2} \left(\erf(γ\sqrt α) - \sgn(\Re γ)\right) .
\end{align*}
Thus,
\begin{equation*}
  I_2 = iπ e^{αγ^2+iβγ} \biggl(\sgn(\Re γ) - \erf\biggl(\frac{iβ}{2\sqrt α} + γ\sqrt α \biggr) \biggr) .
\end{equation*}
\item By the Sokhotski-Plemelj theorem,
\begin{align*}
  \MoveEqLeft[0] \pint \frac{e^{-α k^2+β k}} {k - q} \dd k \\
  &= \lim_{ϵ\rightarrow 0} \frac{I_2(α,β,-iq+ϵ) + I_2(α,β,-iq-ϵ)} {2} \\
  &= iπ e^{-αq^2+βq} \erf\biggl(iq\sqrt α - \frac{iβ}{2\sqrt α} \biggr) .
\end{align*}
\item Taking $β = ir$ and $q=0$ in (c), the limit $α ↘ 0$ can be taken if the integral is understood in the sense of distributions. This yields
\begin{align*}
      ∫ 𝒫 \frac{e^{irk}}{k} \dd k &= -iπ \sgn(r) \\
  ⇒ \quad 2i ℱ^{-1}\left(𝒫 \frac{1}{k} \right) &= \sgn.
\end{align*}
Consequently,
\begin{align*}
  \intpos e^{ikr} \dd r &= (ℱθ)(k) = \frac{1}{2} ℱ(1 + \sgn)(k) \\
     &= πδ(k) + 𝒫 \frac ik \, .
\end{align*}
\end{enumerate}

%\bibliography{references}
\input{main.bbl}

\end{document}

%% file: preamble_math.tex
\usepackage{nccmath}
\usepackage{mathtools} % load after nccmath
\usepackage{isomath}
\usepackage{relsize}
\usepackage{adjustbox}
\usepackage{newunicodechar}
\usepackage{amssymb}
\usepackage{mathrsfs}
\usepackage{siunitx}
\usepackage{braket}
\usepackage{mleftright}
\mleftright
\DeclareMathAlphabet{\mymathbb}{U}{bbold}{m}{n}

\newunicodechar{α}{\alpha}
\newunicodechar{β}{\beta}
\newunicodechar{γ}{\gamma}
\newunicodechar{δ}{\delta}
\newunicodechar{π}{\pi}
\newunicodechar{σ}{\sigma}
\newunicodechar{τ}{\tau}
\newunicodechar{ρ}{\rho}
\newunicodechar{θ}{\theta}
\newunicodechar{ω}{\omega}
\newunicodechar{ν}{\nu}
\newunicodechar{τ}{\tau}
\newunicodechar{ϵ}{\epsilon}
\newunicodechar{ξ}{\xi}
\newunicodechar{ϕ}{\phi}
\newunicodechar{ψ}{\psi}
\newunicodechar{Ψ}{\Psi}
\newunicodechar{Δ}{\Delta}
\newunicodechar{∂}{\partial}
\newunicodechar{ħ}{\hbar}
\newunicodechar{∫}{\int}
\newunicodechar{ℝ}{\mathbb{R}}
\newunicodechar{ℂ}{\mathbb{C}}
\newunicodechar{𝟙}{\mymathbb{1}}
\newunicodechar{∑}{\sum}
\newunicodechar{∞}{\infty}
\newunicodechar{†}{\dagger}
\newunicodechar{±}{\pm}
\newunicodechar{∈}{\in}
\newunicodechar{→}{\rightarrow}
\newunicodechar{↗}{\nearrow}
\newunicodechar{↘}{\searrow}
\newunicodechar{⇒}{\Rightarrow}
\newunicodechar{≤}{\le}
\newunicodechar{≥}{\ge}
\newunicodechar{∼}{\sim}
\newunicodechar{≈}{\approx}
\newunicodechar{≃}{\simeq}
\newunicodechar{≠}{\ne}
\newunicodechar{ℋ}{\mathscr{H}}
\newunicodechar{ℱ}{\mathscr{F}}
\newunicodechar{𝒫}{\mathscr{P}}
\newunicodechar{𝒪}{\mathscr{O}}
\newunicodechar{𝐫}{{\vec r}}
\newunicodechar{𝐱}{{\vec x}}
\newunicodechar{𝐲}{{\vec y}}
\newunicodechar{…}{\dots}
\newunicodechar{⋯}{\cdots}
\newunicodechar{⋅}{\cdot}
\newunicodechar{²}{^2}
\newunicodechar{³}{^3}

\newcommand*\conj[1]{\widebar{#1}}
\newcommand*\smallrel[1]{\mkern1.5mu{#1}\mkern1.5mu}
\renewcommand*\vec[1]{\mathbfit{#1}}
\newcommand*\divslash{}
\AtBeginDocument{\renewcommand*\divslash{\mathbin{/}}}
\newcommand*\dd[1]{\mathop{d#1}}

\newcommand*\intinf{\int_{\mathbb{R}}}
\newcommand*\intpos{\int_{\mathbb{R}_{\mathrlap{+}}}}
\newcommand*\pint{𝒫\mathord{\intinf}}

\DeclareMathOperator\sgn{sgn}
\DeclareMathOperator\erf{erf}
\DeclareMathOperator\erfcx{erfcx}
\DeclareMathOperator\Li{Li}
\DeclareMathOperator\Tr{Tr}
\AtBeginDocument{
  \let\Re\relax
  \let\Im\relax
  \DeclareMathOperator\Re{Re}
  \DeclareMathOperator\Im{Im}
}
\DeclarePairedDelimiter\abs{\lvert}{\rvert}
\DeclarePairedDelimiter\absq{\lvert}{\rvert^2}

\newcommand*\vol{\Omega}

\newcommand*\bound{\text{b}}
\newcommand*\cont{\text{c}}

\newcommand*\BB{\text{\textsmaller{BB}}}

\newcommand*\Aop{\hat{A}}
\newcommand*\mB{m}
\newcommand*\VIB{V^{\aIB}}
\newcommand*\aIB{a}
\newcommand*\aBB{a_\BB}
\newcommand*\HIB{H^{\aIB}}
\newcommand*\Hfree{H^{0}}
\newcommand*\HmanyIB{ℋ^{\aIB}}
\newcommand*\Hmanyfree{ℋ^{0}}
\newcommand*\NB{N}
\newcommand*\nB{n}
\newcommand*\Tc{T_\mathrm{c}}
\newcommand*\mode{\psi^{\aIB}}
\newcommand*\freemode{\psi^{0}}
\newcommand*\projector[1]{#1 \otimes \conj{#1}}

\newcommand*\halfhphantom[1]{\adjustbox{scale=0.5}{\hphantom{#1}}}
\newenvironment{annotate}[1][8cm] {
  \halfhphantom{${}={}$}
  \begin{minipage}{#1}
  \footnotesize
  
  \begin{tabular}{ | p{\columnwidth - 1em} @{} }
} {
  \end{tabular}\end{minipage}
}